\documentclass{article}

\usepackage{arxiv}

\usepackage[utf8]{inputenc} 
\usepackage[T1]{fontenc}    
\usepackage{hyperref}       
\usepackage{url}            
\usepackage{booktabs}       
\usepackage{amsfonts}       
\usepackage{nicefrac}       
\usepackage{microtype}      
\usepackage{graphicx}
\usepackage{doi}

\newcommand{\vrkit}{{\sc IndoorKit}}
\newcommand{\omni}{{\sc Omniverse}}
\newcommand{\source}{https://github.com/realvcla/VRKitchen2.0-Tutorial}
\newcommand{\tutorial}{https://vrkitchen20-tutorial.readthedocs.io/en/latest/}

\title{VRKitchen2.0-\vrkit: A Tutorial for Augmented Indoor Scene Building in Omniverse}

\author{ Yizhou Zhao$^*$ \\
	Department of Statistics\\
	University of California, Los Angeles\\
	\texttt{yizhouzhao@g.ucla.edu} \\
	\And
	Steven Gong$^*$ \\
	Department of Computer Science\\
	University of California, Los Angeles\\
	\texttt{nikepupu@g.ucla.edu} \\
	\And
	Xiaofeng Gao \\
	Department of Statistics\\
	University of California, Los Angeles\\
	\texttt{schaffergao@gmail.com} \\
	\And
	Wensi Ai \\
	Department of Computer Science\\
	University of California, Los Angeles\\
	\texttt{va0817@g.ucla.edu} \\
	\And
	Song-Chun Zhu \\
	Department of Statistics\\
	University of California, Los Angeles\\
	\texttt{sczhu@stat.ucla.edu} \\
}

\renewcommand{\shorttitle}{\textit{arXiv} Template}

\begin{document}

\def\thefootnote{*}\footnotetext{These authors contributed equally to this work}\def\thefootnote{\arabic{footnote}}

\maketitle

\begin{abstract}
With the recent progress of simulations by 3D modeling software and game engines, many researchers have focused on Embodied AI tasks in the virtual environment. However, the research community lacks a platform that can easily serve both indoor scene synthesis and model benchmarking with various algorithms. Meanwhile, computer graphics-related tasks need a toolkit for implementing advanced synthesizing techniques. To facilitate the study of indoor scene building methods and their potential robotics applications, we introduce \vrkit: a built-in toolkit for \textsc{Nvidia} \omni\ that provides flexible pipelines for indoor scene building, scene randomizing, and animation controls. Besides, combining Python coding in the animation software \vrkit\ assists researchers in creating real-time training and controlling for avatars and robotics. The source code for this toolkit is available at \url{\source}, and the tutorial along with the toolkit is available at \url{\tutorial}.
\end{abstract}


\section{Introduction}

Simulation engines are tools used by designers to code and plan out a simulation environment quickly and easily without building one from the ground up. As the development of those simulation engines, researchers and game designers are deploying the recent advances in the field of artificial intelligence (AI) to train autonomous and intelligent agents which grow from experimental laboratories into executable products~\cite{onosato1993development}.

However, even though learning-based algorithms have gradually increased their influence on training agents in virtual environments, there is still a lack of the toolkit that connects the simulation environment and the state-of-the-art developments in the AI community, including innovative tasks, comprehensive datasets, and powerful algorithms~\cite{park2020developing}.

Recently, with the recent release of \textsc{NVIDIA Omniverse}\footnote{\url{https://developer.nvidia.com/nvidia-omniverse}}, a scalable development platform for simulation and design collaboration, researchers can deploy recent advances in AI in \omni\ due to its indispensable features:
\begin{itemize}
    \item Python has taken a lead in determining the programming language for AI or neural networks, and \omni\  exposes much of its functionality through Python bindings;
    \item In \omni\,  both physics simulation and the neural network policy training reside on GPU and communicate directly~\cite{makoviychuk2021isaac};
    \item The universal scene description (USD) format in \omni\ contains many details in 3D computer graphics scene elements and is supported by a wide range of 3D modeling software (e.g. Blender and Autodesk Maya) and game engines (e.g. Unreal Engine and Unity). 
\end{itemize}

\begin{figure}[t]
    \centering
    \includegraphics[width=\textwidth]{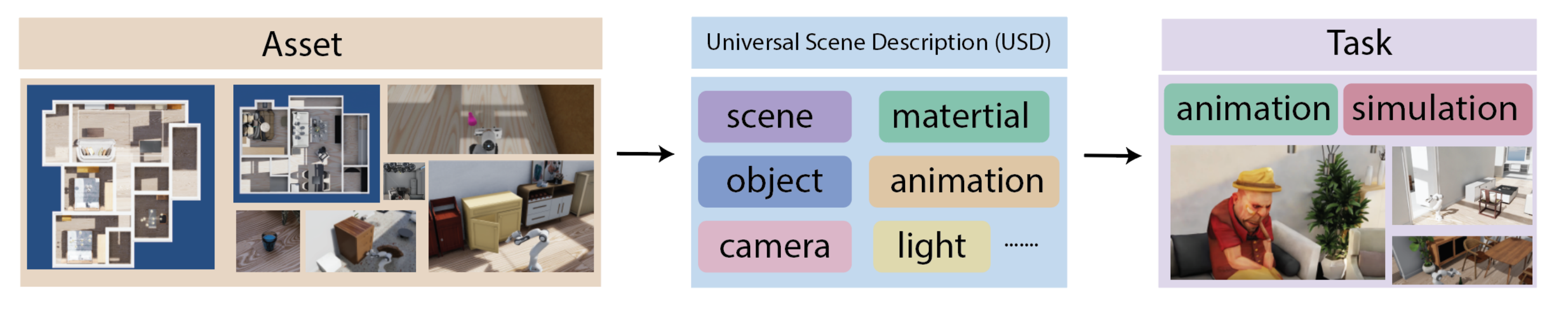}
    \caption{\textbf{Toolkit overview}. We present this tutorial for our new toolkit \vrkit. (1) \vrkit\ supports a wide range of indoor scene datasets synthesized or designed manually, and it allows users to set up their custom dataset. (2) We provide the data processing modules to store scenes that may be from various types of formats into the USD format, making them transferable for other 3D software of game engines. (3)~\vrkit\ can be connected with machine learning models for downstream tasks. (Tasks including but not limited to character animation, physical simulation, and robotics.}
    \label{fig:my_label}
\end{figure}

We present \vrkit: a toolkit built in \textsc{Nvidia} \omni\ that provides flexible pipelines for scene building, character animation, and robotic controls. The innovative features of our \vrkit\ include but not limited to:

\begin{itemize}
    \item Photo-realistic scene rendering by utilizing a wide range of popular 3D assets (e.g. 3D-Front~\cite{fu20213d}, iThor~\cite{kolve2017ai2}, SAPIEN~\cite{xiang2020sapien}, AKB-48~\cite{liu2022akb}, e.t.c.);
    \item Real-time character and robot control by leveraging 3D animatable assets(e.g. AMASS~\cite{mahmood2019amass}, Adobe Mixamo~\cite{adobe2020mixamo};
    \item Comprehensive and flexible pipeline for data labeling, model training testing.
\end{itemize}

Besides, \vrkit\ is an \textit{early work} related to \omni\ and we hope that this work paves the way for the creation of influential representation learning in the future. We also provide extensive documentation including a massive amount of demos and tutorials to encourage related research.

\section{Library overview}

\vrkit\ is developed by the Center for Vision, Cognition, Learning, and Autonomy at the University of California, Los Angeles. It enables easy data loading and experiment sharing for synthesizing scenes and animation with the Python API in \omni. This section briefly describes several features of our \vrkit. \\ 

\subsection*{Working with 3D scene assets} 

We integrate multiple indoor scene datasets in \vrkit. For datasets with different parameterizations of the body, we include documents for meta-data descriptions and visualization tools to illustrate the characteristics of each dataset. Datasets covered include, but not limited to, 3D-Front~\cite{fu20213d}, scenes of AI2Thor~\cite{kolve2017ai2}, and other indoor scene building pipelines~\cite{mahmood2019amass}. Besides, to facilitate a contribution to the community, \vrkit\ also provides detailed instructions for users to upload and pre-process their custom datasets.

\subsection*{Why we choose Omniverse}

There are several reasons for us to choose the newly platform: \omni.

First, for its powerful simulation support: rigid body, soft body, articulated body, and fluid are the main types of simulation that are supported in \omni. 
\begin{itemize}
    \item Rigid body: This type simulates the physics in a static environment. Users can set your mass and gravity for each object and assign forces to objects such as friction or buoyancy.
    \item Soft body: based on the concept of Soft Body Dynamics (SBD), the soft body is controlled by mathematical equations which describe how physical objects behave in the real world when they collide with other objects or themselves. The problem with traditional computer graphics is that the physics simulation can only be done by hand and it takes a lot of time to get an accurate result. With \omni: soft body this task can be done automatically.
    \item Articulation body: working with \omni, \vrkit~provides the tool that enables us to build physics articulations such as robotic arms, kinematic chains, and avatars that are hierarchically organized. It also helps us get realistic physics behaviors in the context of simulation for industrial applications.
    \item Fluid: \omni~allows you to simulate the behavior of liquids and gases. It's based on the Navier-Stokes equations, which describe how fluids flow in an environment. \vrkit~connects the set up of fluid with indoor scene assets, which provides better flexibility of custom tasks.
\end{itemize}

Second, for its Python scripting environment, it is easy to bring open-source and third-party Python libraries into \omni\ to help the research.

\begin{itemize}
    \item  Python is a general-purpose programming language that can be used for many different tasks. It has a simple syntax and readable code, and it’s easy to learn. Python is also very popular in the industry, because of its high performance, robustness, and versatility.
    \item Deep learning can be applied to many different fields from computer vision, speech recognition, natural language processing (NLP), translation, robotics, and much more. Deep learning with Python using Pytorch~\cite{stevens2020deep} and Tensorflow~\cite{dillon2017tensorflow} libraries can be easily supported in \omni.
    
\end{itemize}

Third, \omni~has the powerful rendering ability with the ray tracing technology.
\begin{itemize}
    \item Ray tracing is a system that improves the lighting in the simulation. It cab be used in everything from reflections to shadows both in the environment and scene elements including atmospheric effects, surfaces reflections and even diffused lighting. 
    \item Besides, users can render the scenes with the universal scene description (USD) format. This is a highly compressed format that allows to store the entire scene information in a single file. It is also very simple to use and understand. The basic idea behind USD is that it stores all of the information about how the scene is rendered, including models, textures, lights, cameras, and animation.
\end{itemize}

\section{Potential use case}

Another essential goal of \vrkit\ is to apply the state-of-the-art embodied AI research while synthesizing photo-realistic and physical-realistic rendering. Here, we list a few common use cases of our library. The full demo and tutorial for them can be found at \url{\source}\\

\begin{figure}[t]
    \centering
    \includegraphics[width=\textwidth]{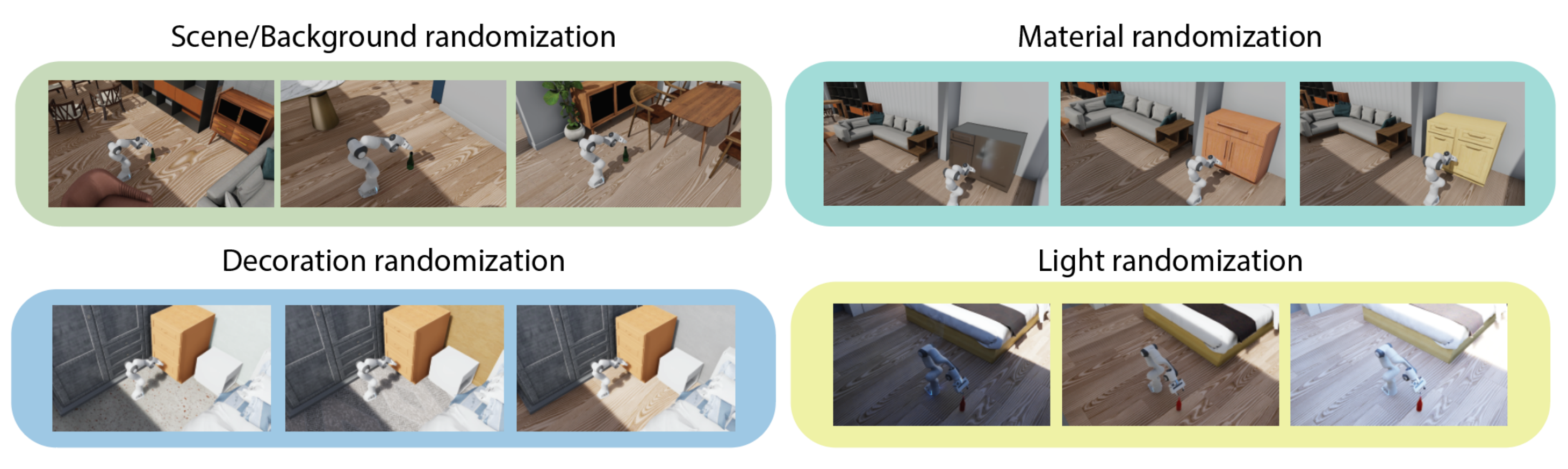}
    \caption{\textbf{We also provide different randomization strategies for indoor scenes}. (1) Scene/Background randomization: the same simulation task can share different backgrounds. (2) Material randomization: scene items can be rendered as randomized materials. (3) Decoration randomization: colors and materials of the wall and floor can vary. (4) Light randomization: simulation tasks can be performed under different lighting conditions.}
    \label{fig:example}
\end{figure}

\subsection*{Indoor scene labeling}

Indoor scene labeling is the process of identifying and labelling indoor scenes. The goal of this project is to improve the quality, quantity and consistency of indoor scene labels in an effort to facilitate more efficient research use. The field of the EAI still lacks of meaningful and near-realistic indoor scenes. We apply \omni~ platform to perform labeling steps for indoor scenes. The sampled data piece contains the information of the game, and allows users to perform downstream tasks. \vrkit\ also offers a clean interface for labeling scene and robot information.

\subsection*{Indoor character animation}

With the recent progress in 3D character animation creation, the popularity of animation generation, as well as its application, grows. Meanwhile, with the improvement of virtual environment simulation capability, researchers have started to study agent behavior in the context of Embodied AI. However, the production process of adapting the generated animation in a photo-realistic and physics-reliable environment is laborious. The key idea is to make a balance between the original animation clip, physics scene, and social interaction meaning through reinforcement learning on the kinematics. 

\subsection*{Robotics oriented simulation}

 Robotics is a field of study that involves the design, construction and operation of robots. The goal of robotics in \omni\ is to build machines capable of performing tasks that are difficult or impossible for humans to perform. Robotics can be used in many different fields including medicine, manufacturing, research and education.

\section{Development and maintenance}

The project is developed by the 
 Center for Vision, Cognition, Learning, and Autonomy at University of California, Los Angeles. \vrkit\ is developed publicly through \textit{Github} with an issue tracker to report bugs and ask questions. Documentation consists 
of tutorials, examples, and API documentation.
The third-party packages include Pytorch~\cite{paszke2019pytorch} for the deep learning framework, and Jupyter notebook~\cite{randles2017using}. For the demo and tutorial, please visit \url{\tutorial}..

\section{Conclusion}
We presented \vrkit, an Python library to help researchers to easily develop their indoor scene synthesizing methods and apply the scene to EAI tasks.

\bibliographystyle{unsrt}
\bibliography{citations}

\end{document}